\documentclass[11pt,twoside]{article}
\usepackage{./asp2021}

\aspSuppressVolSlug
\resetcounters

\begin{document}

\title{Deep and low mass-ratio contact binaries and their third bodies}
\author{Zhu Liying,$^{1,2,3}$ Qian Shengbang,$^4$ Liao Wenping,$^{1,2,3}$ Zhang Jia,$^{1,2,3}$ Shi Xiangdong,$^{1,2,3}$ Li Linjia,$^{1,2,3}$ Meng Fangbin,$^{1,2,3}$ Wang Jiangjiao$^{1,2,3}$ Matekov Azizbek $^{1,2,5}$}
\affil{$^1$ Yunnan Observatories, Chinese Academy of Sciences, Kunming 650216, China; \email{zhuly@ynao.ac.cn}}
\affil{$^2$ Key Laboratory of the Structure and Evolution of Celestial Objects, Chinese Academy of Sciences, Kunming 650216, China}
\affil{$^3$ University of Chinese Academy of Sciences, No.1 Yanqihu East Rd, Huairou District, Beijing, P.R.China 101408}
\affil{$^4$ School of Physics and Astronomy, Yunnan University, Kunming 650091, China}
\affil{$^5$ Ulugh Beg Astronomical Institute, Uzbekistan Academy of Sciences, 33 Astronomicheskaya str., Tashkent, 100052, Uzbekistan}

% This section is for ADS Processing.  There must be one line per author.
\paperauthor{Zhu, L. -Y.}{zhuly@ynao.ac.cn}{}{Chinese Academy of Sciences}{Yunnan Observatories}{Kunming}{Yunnan Province}{650216}{China}
\paperauthor{Qian, S. -B.}{qsb@ynao.ac.cn}{}{Yunnan University}{School of Physics and Astronomy}{Kunming}{Yunnan Province}{650091}{China}
\paperauthor{Liao, W. -P.}{liaowp@ynao.ac.cn}{}{Chinese Academy of Sciences}{Yunnan Observatories}{Kunming}{Yunnan Province}{650216}{China}
\paperauthor{Zhang J.}{zhangjia@ynao.ac.cn}{}{Yunnan University}{School of Physics and Astronomy}{Kunming}{Yunnan Province}{650091}{China}
\paperauthor{Shi X.-D.}{sxd@ynao.ac.cn}{}{Chinese Academy of Sciences}{Yunnan Observatories}{Kunming}{Yunnan Province}{650216}{China}
\paperauthor{Li L.-J.}{lipk@ynao.ac.cn}{}{Chinese Academy of Sciences}{Yunnan Observatories}{Kunming}{Yunnan Province}{650216}{China}
\paperauthor{Meng, F.-B.}{mengfangbin@ynao.ac.cn}{}{Chinese Academy of Sciences}{Yunnan Observatories}{Kunming}{Yunnan Province}{650216}{China}
\paperauthor{Wang, J.-J.}{wangjiangjiao@ynao.ac.cn}{}{Chinese Academy of Sciences}{Yunnan Observatories}{Kunming}{Yunnan Province}{650216}{China}

\begin{abstract}
Deep and low mass-ratio contact binaries (DLMCBs) are believed to be in the final stage of their contact phase, potentially leading to the formation of fast-rotating single stars such as FK Com-type stars and blue stragglers, as well as luminous red novae. These systems serve as an excellent laboratory for studying stellar coalescence and merging processes. Our search for DLMCBs began in 2004 and has since identified a group of such systems. Together with that collected from the literature, more than 100 DLMCBs have been detected so far. Half of them have had their periods investigated based on O-C curves. Some have shown period increases, while others have exhibited period decreases. Among them, more than half DLMCBs have cyclic variations, suggesting the possibility of the existence of a third body orbiting around the DLMCBs. Furthermore, with more data obtained extending the span of the O-C curve, more cyclic variations could be detected. The high proportion of signs of the presence of third bodies makes them an essential factor to consider when studying the merger of contact binaries.
\end{abstract}

\section{Introduction}
Contact binaries comprise two stars with components intricately connected, sharing a common convective envelope (CCE). This shared envelope leads to the exhibition of EW-type light curves with similar depths in both minima, even when the two components differ significantly in mass. Contact binaries can originate from various processes, including the fragmentation of a molecular cloud, the capture of a passing star, and stellar collisions. Over time, a combination of mechanisms such as tidal friction, magnetic braking, Kozai cycles within triple systems, and mass transfer between the components acts to bring the stars into closer proximity, giving rise to contact binaries. This continuous evolution may eventually lead to the merger of the two stars. The well-known luminous red nova, V1309 Scorpii, provides confirmed evidence that contact binaries can indeed undergo the merger process.

A luminous red nova is a stellar explosion characterised by a distinct red colour, and a light curve that fades slowly with resurgent brightness in the infrared. They were not only discovered in the Milky Way galaxy such as V4332 Sgr (Martini et al. 1999), V838 Mon (Munari et al. 2002; Bond et al. 2003) and V1309 Sco (Mason et al. 2010), but also discovered in other Galaxies including the nearest major galaxy M31(Mould et al. 1990) and the elliptical galaxy M85 (Kulkarni et al. 2007). V1309 Sco was discovered in September 2008 (Nakano 2008) and underwent an evolutionary transition from an F-type giant to a late M-type giant (Mason et al. 2010). it presented a unique opportunity for the study of stellar mergers, facilitated by the recorded light curves of its progenitor captured by OGLE. These light curves strongly suggested that V1309 Scorpii originated as a contact binary (Tylenda et al. 2011). Zhu et al. (2016) further revealed that its progenitor was a low-mass ratio (q~0.1) and deep contact binary star (f=90\%) in 2002. The search for and study of deep and low mass-ratio contact binaries (DLMCBs) holds significance in unraveling the destiny of contact binaries and advancing our understanding of stellar coalescence and merging processes.

The criterion for binary star merging is defined by the condition that the orbital angular momentum is less than three times the rotational angular momentum (Hut, 1980). Given the orbital constraints imposed by the contact configuration, contact binaries with lower mass ratios tend to approach this merging condition. On the other hand, deep contact implies a thick common envelope (CE), which is inherently unstable. Hence, contact binaries characterized by both deep contact and low mass ratios emerge as promising objects for the study of stellar mergers. Our exploration of deep and low mass-ratio contact binaries (DLMCBs) began in 2004 (Qian \& Yang 2004), and we proposed that a contact binary could be categorized as a DLMCB if its mass ratio $q \leq 0.25$ and the fill-out factor $f \geq 50\%$ (Qian et al., 2005a; 2006b). This classification provides a useful framework for identifying binaries that are conducive to investigating the dynamics of stellar merging.

\section{Candidates selection}
Thanks to several photometric surveys conducted around the world, a significant number of eclipsing binaries with EW-type light variation have been discovered.Together with spectroscopic survey by the Large Sky Area Multiobject Fiber Spectroscopic Telescope (LAMOST), not only the light curves, but also the stellar atmospheric parameters including the effect temperature $T_{eff}$ , the gravitational acceleration Log(g), the metallicity [Fe/H] and the radial velocity $V_{r}$ of a large number of contact binaries have been obtained. This provides a valuable database for studying the physical properties of contact binaries. By cross-checking the VSX catalog (the international variable star index, Watson et al. 2006) with LAMOST targets, we have compiled a catalog of EW binaries with all the available parameters and continue to update it with new observations (Qian et al., 2017, 2020). The entire catalogue is available online, with an electronic version accessible through the website (http://search.vbscn.com/2020EW.table1.txt).

Based on the catalog, a new orbital period distribution of contact binaries has been derived with a strong maximum and a very sharp edge at around 0.15 days. The maximum of the distribution is at about 0.31 d, and most EWs are in the orbital period range from 0.285 to 0.345 d (Qian et al., 2020). The period-color (or temperature) relation is a well-known relation for contact binaries (Eggen 1967; Rucinski 1998). To investigate this relation in detail using LAMOST stellar atmospheric parameters, the new relation is presented in Figure 1. As shown in the figure, most EWs are located within the two cyan lines, which represent the boundaries of normal EWs. Green dots represent EWs observed by low-resolution spectra (LRS), while dark dots represent EWs observed by medium-resolution spectra (MRS). Systems near the right boundary usually have longer orbital periods for a given temperature and higher orbital angular momentum. They usually have marginal (or shallow) contact configurations with fill-out factors less than 20\%. Systems located near the left boundary have shorter orbital periods and are usually deep contact systems. Therefore, the LAMOST data are very useful for selecting targets for detailed follow-up observations and investigations, and more and more DLMCBs will be detected in the future.

\articlefigure[width=0.85\textwidth]{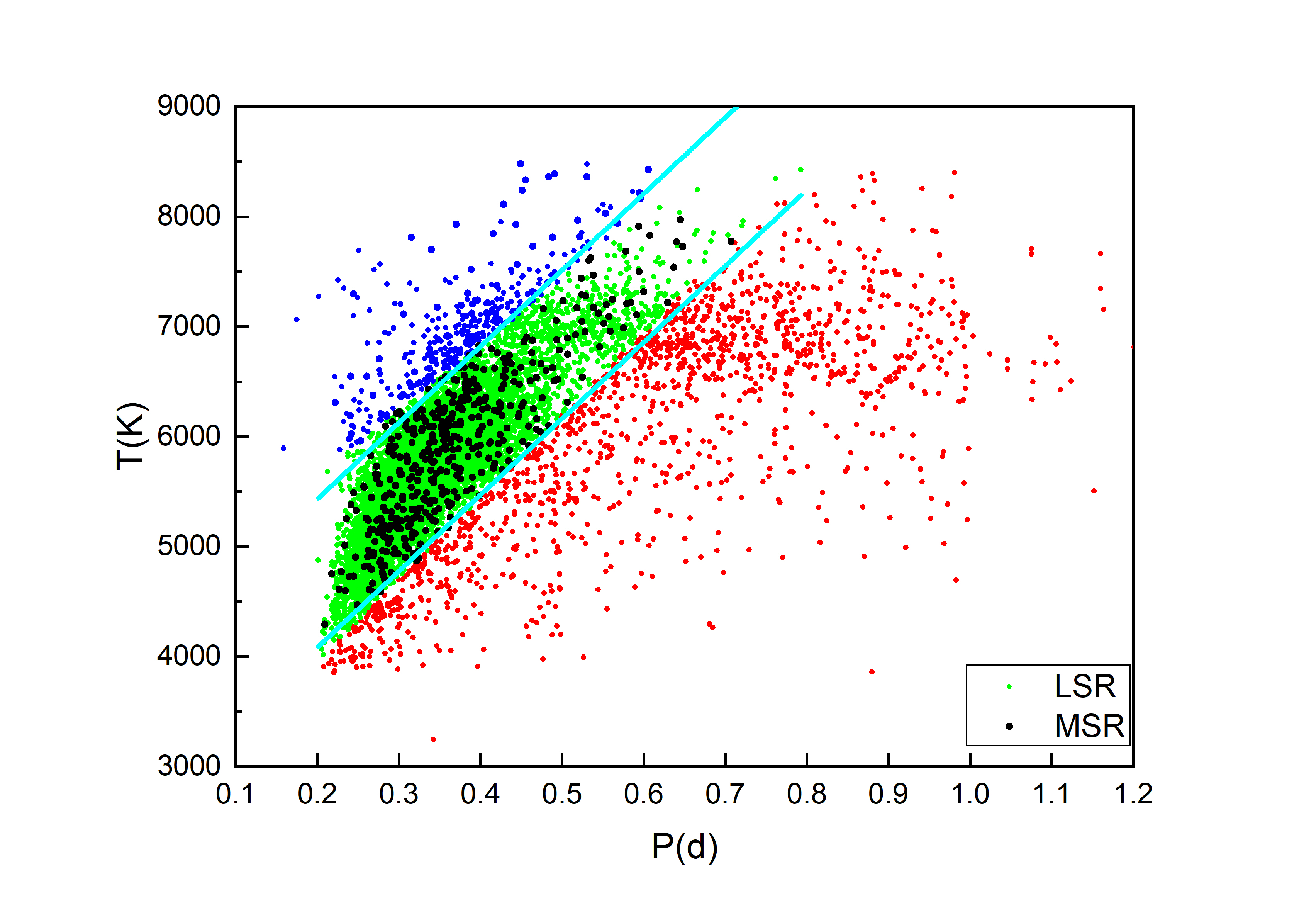}{fig1}{Correlation between the orbital period and effective temperature based on normal EWs observed by LRS and MRS (green and dark dots respectively). Blue dots refer to binaries located above the left boundary of normal EWs, while red dots to systems below the right boundary. Systems near the left border are deep contact binaries.}

\section{Recent results}
Since 2004, our team has published a series of papers to report the discovery of Deep and Low Mass-Ratio Contact Binaries (DLMCBs) (Qian \& Yang, 2004; Qian et al., 2005a, 2005b, 2006, 2007, 2011; Zhu et al., 2005, 2011; Yang et al., 2005, 2009, 2013; Liao et al., 2017, 2022; Sarotsakulchai et al., 2018; Meng et al., 2023). In addition to our own discoveries, many DLMCBs have been reported by other researchers, including Pribulla et al., 2009; Zola et al., 2017; Kjurkchieva et al., 2017; ; Li et al., 2022; Christopoulou et al., 2022; Guo et al., 2023 etc. To date, more than 100 DLMCBs have been discovered, and approximately half of them have had their periods investigated based on the O-C curve. Among these systems, 33 show period increases, 20 show period decreases, and 25 exhibit periodic variations due to the light travel time caused by a third body orbiting around the central DLMCBs. Here we introduce our recent results on two active DLMCBs with confirmed additional body based on spectroscopic observation and the changes of the O-C curve due to the light travel time effect.

\subsection{NY Bootes: An Active DLMCB with a Cool Companion in a Hierarchical Triple System}
NY Bootes (NY Boo, TIC 309809572) was first discovered as an overcontact binary system by the Northern Sky Variability Survey (Hoffman et al. 2009). The Transiting Exoplanet Survey Satellite (TESS; Ricker et al. 2015) provided high-precision photometric data for NY Boo in five sectors (sector16, sector23, sector24, sector49, and sector50), with an exposure time of 2 minutes. The phased light curves for all sectors are shown in the upper-left panel of Figure 2. As can be seen from this figure, the light curves of NY Boo exhibit noticeable variations over time, indicating significant magnetic activity in the system. Six low-resolution spectra of NY Boo are available from the LAMOST database, and three infrared spectra were obtained from the SDSS Scientific Archive Server. By using the cross correlation function (CCF) method on the LAMOST and SDSS spectra, we obtained its radial velocity (RV) curve, which is shown in the upper-right panel of Figure 2. A simultaneous analysis of the light curve and RV curve reveals that NY Boo is a contact binary with a mass ratio ($q = 0.139$) and fillout factor ($f = 73\%$). It is worth noting that NY Boo is a partial eclipsing binary with an inclination of only 67.4 degrees, and its light curves exhibit significant changes in a short time. This makes it difficult to confirm that it is a DLMCB without the support of RV curve. Most of the DLMCBs detected to date are total eclipsing binaries, as the reliable photometric mass ratio can be determined in this case. LAMOST is a reflecting Schmidt telescope with an effective aperture of 4 m that can simultaneously observe and obtain spectra for close to 4000 targets. With more spectra of EW binaries, more DLMCBs could potentially be discovered in the future.

In order to study the orbital period variation, we collected all available timings of NY Boo, together with the others calculated by us, we constructed the O-C curve of NY Boo based on these times of light minimum, which span 23 years. A detailed investigation of its O-C diagram reveals that NY Boo undergoes a long-term period decrease while also experiencing cyclical variations in its orbital period. The corresponding fit curves can be seen in the lower-left panel of Figure 2. The rate of orbital period decrease is $dP/dt = -1.305( \pm 0.022) \times 10^{-6} day yr^{-1}$. Comparing the orbital period reduction timescale with the thermal timescale of the primary component, we found that NY Boo may be in the fast-shrinking stage. The cyclical variation of the orbital period with an amplitude $A = 0.00713$ day and a period $P = 16.9 yr$ could be due to the light-travel time effect of the tertiary companion, which is consistent with the third light features detected from the CCF profiles of the SDSS spectra (lower-right panel of Figure 2). Therefore, NY Boo should be a triple system. Evidence of the existence of the third body was only found in the infrared spectrum, suggesting that it is a late-type star. By combining the third light contribution derived from the CCF fit with the results of the O-C curve fit, we derived the orbital inclination, mass, and semimajor axis of the third body as $i_{3} = 50°$, $m_{3} = 0.31 M_{\bigodot}$, and $a_{3} = 5.82$ au, respectively, This indicates that the third body is non-coplanar with the central pair. The combination of a low mass ratio, high fill-out factor, fast-shrinking, and rapid light variation in the inner pair with a non-coplanar tertiary component makes NY Boo an excellent target for studying the late evolution of contact binaries and stellar mergers (Meng et al., 2023).

\articlefigure[width=0.98\textwidth]{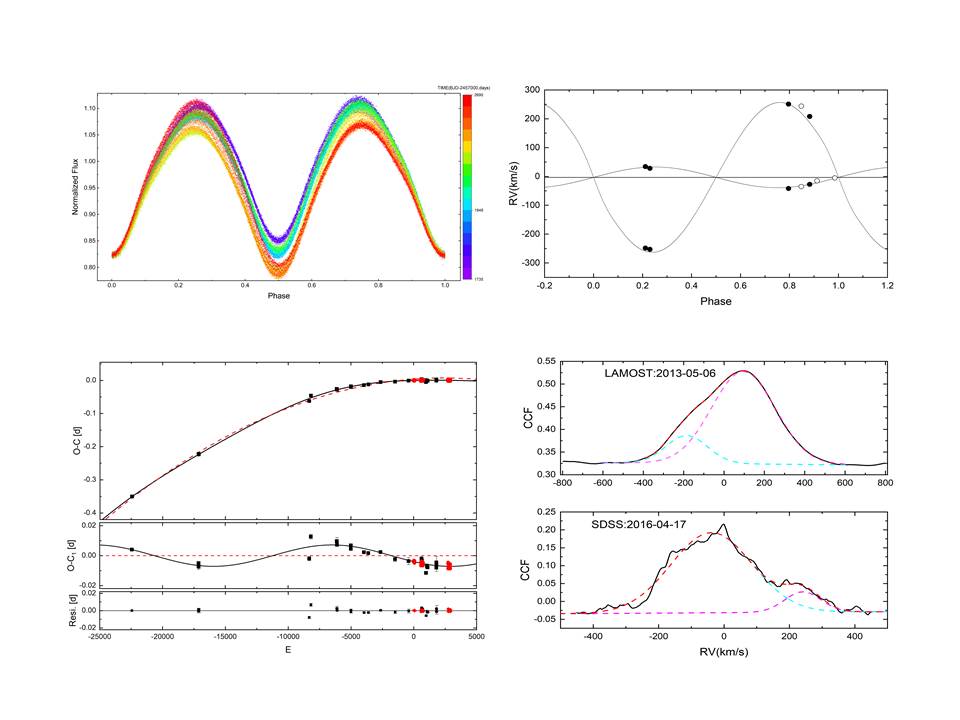}{fig2}{Upper-left panel: The TESS phase-flux diagram of NY Boo that varied with time. Upper-right panel:The RV curves of NY Boo. The filled dots represent data from LAMOST spectra and open dots represent data from SDSS spectra. Lower-left panel: The O-C diagram of NY Boo. The red dots refer to the eclipse times obtained from TESS. Lower-right panel: CCF profile of the LAMOST (upper panel) and SDSS (lower panel) example spectra. The solid black line indicates the CCF profile. The red dashed line represents the double Gaussian fit. The blue and pink dashed lines denote individual Gaussian components.}

\subsection{V410 Aur: An Active DLMCB in a quintuple stellar system}

The variability of V410 Aur was first discovered by the Hipparcos satellite, and its radial velocity study was conducted by Rucinski et al.(2003). They determined that V410 Aur is a spectroscopic double-lined binary with a mass ratio of $q_{sp} = 0.144(13)$ and third light contribution $l_{3} = 26(1)\%$. The light curves of V410 Aur are known to vary with time, indicating the star's activity. In 2005, we modeled the light curves of V410 Aur and concluded that it is a DLMCB (Yang et al. 2005). However, the study of the O-C curve at that time suggested that the orbital period of V410 Aur was continuously increasing.

In 2022, we reanalyzed the O-C curve and found that the previous period increase was only part of the cyclic variation (Liao et al., 2022). We analyzed the cyclical variation for the light-travel time effect and determined that the minimum mass of the third body is $1.39 (±0.13) M_{\bigodot}$ which is much larger than the spectroscopically inferred value of (0.97 $M_{\bigodot}$ ). This indicates that the spectroscopically detected tertiary is actually a single-lined spectroscopic binary with an unseen component. We also determined the maximum orbital semimajor axis of the third body to be $6.19 (±0.67)$ au. Gaia detected a visual companion to V410 Aur at practically the same distance from the Sun, providing further confirmation of its physical bond. These results reveal that V410 Aur contains a single-lined spectroscopic binary with a visual companion in a quintuple stellar system, making it an interesting active merger progenitor in a quintuple system (2 + 2 + 1).

\section{Summary}
We have gathered the periods of the third bodies in DLMCBs systems and plotted their relationship with the period of the DLMCBs in Figure 3. As seen from this figure, longer period DLMCBs tend to have farther third bodies. The periods of the third bodies vary from several years to decades, with most having a period greater than 10 years. This suggests that it typically takes more than 10 years of data coverage to detect their third bodies. Most newly discovered DLMCBs do not have long observational history, which makes it difficult to detect the third body based on the light-travel time effect. Additionally, many DLMCBs exhibit variable light curves, which are often attributed to the magnetic activity of the binary component. In such cases, the cyclic changes detected in the O-C curve may be explained by magnetic activity cycles (Applegate 1992). However, both NY Boo and V410 Aur consist of active stars, and their periodic changes detected in the O-C curves are confirmed to be due to the third body through their spectral features. Therefore, the presence rate of additional companions in DLMCB systems may be underestimated.

In addition to the O-C curve, the presence of a third body can also be detected through spectroscopic data, as seen in the cases of NY Boo and V410 Aur. Since the period of DLMCBs is much shorter than that of the third body, the spectral lines of the binary components will exhibit blue or red shift due to the Doppler effect, while the spectral lines of the third body remain relatively motionless. LAMOST has initiated a time-domain spectroscopic survey, which will observe each of the 200,000 stars an average of 60 times. This data will allow for the detection of additional bodies orbiting around DLMCBs in various forms in the future. The high proportion of third bodies present indicates that they will be an essential factor to consider when studying the merger of contact binaries.

\articlefigure[width=0.85\textwidth]{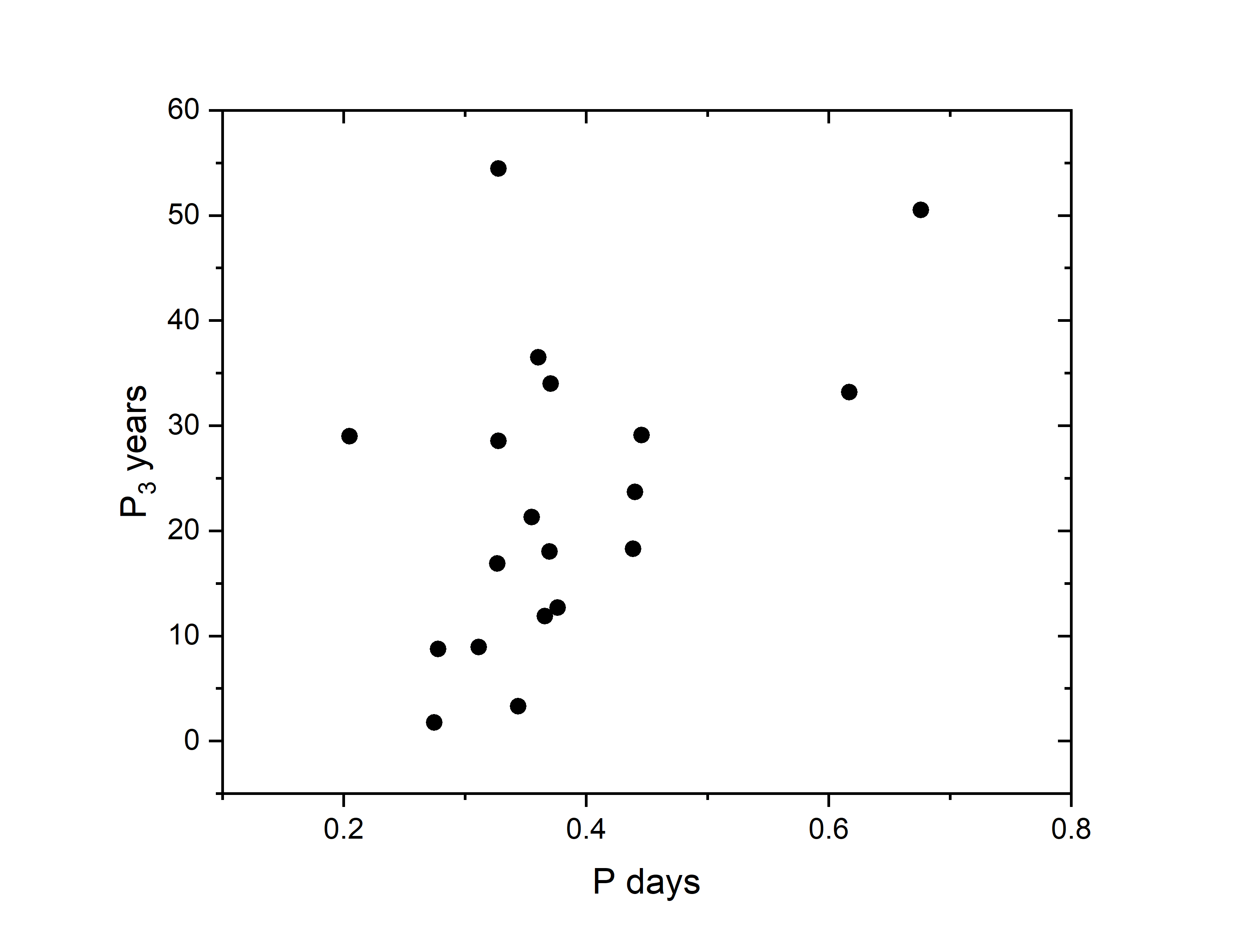}{fig3}{The relation between the period of DLMCBs and the periods of the third bodies. }

\acknowledgements This work is supported by the International Cooperation Projects of the National Key R\&D Program (No. 2022YFE0127300), the Young Talent Project of "Yunnan Revitalization Talent Support Program" in Yunnan Province and CAS "Light of West China" Program. We are grateful to all staff members of ASP for their support.

%\bibliography{editor}  % For BibTex

% For non-BibTex:

\end{document}